\documentclass[10pt,a4paper]{article}

\usepackage{amsmath}
\usepackage{amssymb}
\usepackage{amsfonts}
\usepackage{listings,color}

\usepackage{fixmath,amsthm}

\usepackage{fixmath}
\newcommand{\vect}[1]{\ensuremath{ \mathbold #1 } }

\newcommand{\R}{{\mathbb R}}

\newcommand{\N}{{\mathbb N}}

\newcommand{\D}[2]{ \ensuremath{ \frac{d #1 }{d #2 } }}

\newcommand{\corr}[1]{#1}

\newtheorem{theorem}{Theorem}[section]

\theoremstyle{definition}
\newtheorem{definition}[theorem]{Definition}

\theoremstyle{remark}

\date{}

\begin{document}
\bibliographystyle{plain}

\title{A Constraint Solving Approach to Tropical Equilibration and Model Reduction}

\author{
Sylvain Soliman$^1$, Fran\c{c}ois Fages$^1$, Ovidiu Radulescu$^2$ \\ \\
\small $^1$ EPI Contraintes, Inria
Paris-Rocquencourt, France\\
\small $^2$ DIMNP UMR CNRS 5235, University of Montpellier 2, Montpellier, France.}

\maketitle

\begin{abstract}
\corr{Model reduction is a central topic in systems biology and dynamical systems theory,
for reducing the complexity of detailed models,
finding important parameters, and developing multi-scale models for instance.
While perturbation theory is a standard mathematical tool to analyze the different time scales
of a dynamical system, and decompose the system accordingly,
tropical methods provide a simple algebraic framework to perform these analyses systematically
in polynomial systems.
The crux of these tropicalization methods is in the computation of tropical equilibrations.
In this paper we show that constraint-based methods, using reified constraints for expressing the equilibration conditions,
make it possible to numerically solve non-linear tropical equilibration problems,
out of reach of standard computation methods.
We illustrate this approach
first with the reduction of simple biochemical mechanisms such as the Michaelis-Menten
and Goldbeter-Koshland models, and second,
with performance figures obtained on a large scale on the model repository \texttt{biomodels.net}.}
\end{abstract}

\section{Preliminaries on Model Reduction by Tropicalization}
We consider networks of biochemical reactions with mass action kinetic laws.
Each reaction is defined as $$\sum_i \vect{\alpha}_{ji} A_i \rightarrow \sum_{k} \beta_{jk} A_k.$$
The stoichiometric vectors  $\vect{\vect{\alpha}_{j}} \in \N^n$, $\vect{\beta_{j}} \in \N^n$ have coordinates
$\vect{\alpha}_{ji}$ and $\beta_{jk}$ and define which species are consumed and produced by the reaction $j$
and in which quantities.

The mass action law  means that reaction rates are monomial functions of the species concentrations $x_i$ and reads
 \begin{equation}
 R_j(\vect{x}) = k_j \vect{x^{\vect{\alpha}_{j}}}.
 \label{rrateone}
 \end{equation}
 where $k_j >0$ are kinetic constants, $\vect{\alpha}_{j} = (\alpha_1^j, \ldots, \alpha_n^j)$ are multi-indices
 and $\vect{x^{\vect{\alpha}_{j}}}  = x_1^{\alpha_1^j} \ldots x_n^{\alpha_n^j}$.

The network dynamics is described by the following differential equations
 \begin{equation}
 \D{x_i}{t} = \sum_j k_j ({\beta_{ji}} - {\alpha_{ji}})  \vect{x}^{\vect{\alpha_{j}}}.
 \label{massaction}
 \end{equation}
In what follows, the kinetic parameters do not have to be known precisely, they are given by their orders
of magnitude. A convenient way to represent orders is by considering that
\begin{equation}
k_j = \bar k_j \epsilon^{\gamma_j},
\end{equation}
where $\epsilon$ is \corr{a} positive parameter much smaller than $1$, \corr{$\gamma_j$ is an integer,
and $\bar k_j$ has order unity.}
An approximate integer order can be obtained from any real positive parameter by
\begin{equation}
\gamma_j = \text{round}( \log(k_j) / \log(\epsilon)),
\end{equation}
where round stands for the closest integer.
For instance,
if $\epsilon = 1/10$,
$\gamma_j$ will represent the \corr{logarithmic} value of the parameter
rounded to the nearest decade. \corr{Notice that in this representation,
small quantities have large orders. Furthermore, the smaller $\epsilon$,
the better the separation between quantities of different orders,
indeed $\lim_{\epsilon \to 0} \frac{k_i}{k_j} = \infty $
if $\gamma_i < \gamma_j$.
We are also interested in the orders of the species concentrations,
therefore we introduce a vector of orders $\vect{a} = (a_1,\ldots,a_n)$, such that
$\vect{x} = \bar{\vect{x}} \epsilon^{\vect{a}}$. Orders $\vect{a}$ are unknown and have to be calculated.
To this aim}, the network dynamics can be described by a rescaled system of ordinary differential equations
 \begin{equation}
 \D{\bar{x}_i}{t} = (\sum_j \epsilon^{\mu_j} k_j ({\beta_{ji}} - {\alpha_{ji}})  {\bar{\vect{x}}}^{\vect{\alpha_{j}}})\epsilon^{-a_i},
 \label{massactionrescaled}
 \end{equation}
where
\begin{equation}
\mu_j = \gamma_j + <\vect{a},\vect{\alpha_j}>,
\label{muj}
\end{equation}
and $<,>$ stands for the vector dot product.
The r.h.s.\ of each equation in
\eqref{massactionrescaled} is a sum of monomials in the concentrations, with positive and negative signs given
by the stoichiometries ${\beta_{ji}} - {\alpha_{ji}}$. Generically, these monomials have different orders (given by $\mu_j$)
and there is one monomial that dominates  the others. In this case, the corresponding
variable will change rapidly in the direction
imposed by this dominating monomial.
However, \corr{on sub-manifolds of the phase space}, at least two monomials, one positive and one negative \corr{may have the same order.
This situation was called tropical equilibration in~\cite{NGVR13arxiv}.
Tropical equilibration is different from equilibrium or steady state in many ways.
Firstly, steady state means equilibration of all species, whereas tropical equilibration may concern only one or a few rapid species. Secondly, steady state means that forces are
rigorously compensated on all variables that are at rest, whereas tropical equilibration means that only the dominant forces are compensated and variables may change slowly under the influence
of uncompensated, weak forces. Compensation of dominant
forces constrains the dynamics of the system to a low dimensional manifold named invariant manifold \cite{RGZN12fbcb,NGVR12sasb}. As discussed in~\cite{NGVR13arxiv}, tropical equilibrations
encompass the notions of quasi-steady state and quasi-equilibrium from singular perturbation theory of
biochemical networks, but are more general.
Let us provide a formal definition of tropical equilibration (see \cite{NGVR13arxiv} for more details).}
\begin{definition} \label{eqreactions}
Two reactions $j$, $j'$ are tropically equilibrated on the species $i$ iff:

i) $\mu_j = \mu_{j'}$,

ii) \corr{
$(\beta_{ji} - \alpha_{ji} ) (\beta_{j'i} - \alpha_{j'i}) < 0$ (meaning that the effects of the reactions $j$ and $j'$ on the species $i$ are opposite)},

iii) $\mu_k \geq \mu_j$ for any reaction $k \neq j,j'$, such that
\corr{
$\beta_{ki} \neq \alpha_{ki} $.
}
\end{definition}

According to \eqref{muj} and Definition~\ref{eqreactions}, the equilibrations correspond to vectors $\vect{a} \in R^n$ where the minimum in the definition of the piecewise-affine function
$f_i(\vect{a}) = \min_j (\gamma_j + <\vect{a},\vect{\alpha_j}>)$ is attained at least twice.
\corr{Tropical equilibrations are used to calculate the unknown orders $\vect{a}$. The solutions have a
geometrical interpretation.}
Let us consider the equality $\mu_j = \mu_{j'}$. This represents the equation of a
$n-1$ dimensional hyperplane of $\R^n$, orthogonal to the vector $\vect{\alpha_j} - \vect{\alpha_{j'}}$:

\begin{equation}
\gamma_j + <\vect{a},\vect{\alpha_j}> = \gamma_{j'} + <\vect{a},\vect{\alpha_{j'}}>
\label{lines}
\end{equation}
For each species $i$, we consider the set of reactions ${\mathcal R}_i$
that act on this species, namely the reaction $k$ is in ${\mathcal R}_i$ iff
$(\vect{\beta_{k}} - \vect{\alpha_{k}})_i \neq 0$.
The finite set ${\mathcal R}_i$ can be characterized by the corresponding set
of stoichiometric vectors $\vect{\alpha_k}$.
The set of points of $\R^n$ where at least two reactions equilibrate on the species
$i$ corresponds to the places where the function $f_i$ is not locally affine (the minimum
in the definition of $f_i$ is attained at least twice). For simplicity, we shall call this
locus tropical  manifold \cite{NGVR13arxiv,Viro08jjm}.


A simple example of biochemical network is the \corr{Michaelis-Menten mechanism of an
enzymatic reaction.} This network consists of
two reactions:
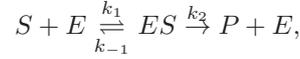
\begin{equation}
S + E \underset{k_{-1}}{ \overset{k_{1}}{\rightleftharpoons}} ES  \overset{k_2}{\rightarrow}
P + E,
\notag
\end{equation}
where $S,E,ES,P$ represent the substrate, the enzyme, the enzyme-substrate complex and the product,
respectively.

The system of polynomial differential equations reads:
\begin{eqnarray}
 x_1' & = -k_1 x_1 x_3 + k_{-1} x_2, \notag \\
 x_2' & = k_1 x_1 x_3 - (k_{-1}+k_2)x_2, \notag \\
 x_3' & = -k_1 x_1 x_3 + (k_{-1} +  k_2) x_2, \notag \\
 x_4' & = k_2 x_2.
\end{eqnarray}
where $x_1=[S]$, $x_2=[SE]$, $x_3=[E]$, $x_4=[P]$.

There are two conservation laws:
$x_2 + x_3 = e_0$ and $x_1 + x_2 + x_4 = s_0$
The rescaled variables are
$x_i= \bar x_i \epsilon^{a_i}$, $1 \leq i \leq 4$,
$k_1= \bar k_1 \epsilon^{\gamma_1}$, $k_{-1}= \bar k_{-1} \epsilon^{\gamma_{-1}}$, $e_0= \bar e_0 \epsilon^{\gamma_e}$, $s_0= \bar s_0 \epsilon^{\gamma_s}$.
Let us notice that the last equation can never
be equilibrated because it contains only one monomial.
\corr{The tropical
equilibration equations for the remaining variables read:}
\begin{eqnarray}
\gamma_{1} + a_1 + a_3 &= \gamma_{-1} + a_2, \notag \\
\gamma_{1} + a_1 + a_3 &= \min(\gamma_{-1}, \gamma_{2}) + a_2, \notag \\
\gamma_{1} + a_1 + a_3 &= \gamma_{2} + a_2, \notag \\
 \min(a_2, a_3) &= \gamma_e, \notag \\
 \min(a_1, a_2, a_4) &= \gamma_s.
\end{eqnarray}
The set of integer orders endowed with the minimum and sum operations
is a semiring, called min-plus algebra \cite{CGQ99arc}
where the minimum is noted $\oplus$ and the sum $\otimes$.
Our tropical equilibration problem is solving a set of polynomial
equations in this semi-ring.

Let us emphasize an important difference between the calculation of
tropical equilibrations and calculation of exact equilibria of systems of
polynomial differential equations. If there are exact conservation laws, the
set of exact equilibrium equations are linearly dependent, therefore one
can eliminate some of them from the system.
Because elements in a min-plus semiring do not generally have additive inverses,
elimination is not automatically possible in systems
of tropical equations. In this case, one should keep all the tropical equilibrium equations
for all the variables and add to them the tropical conservation relations.

\section{Example of Golbeter-Koshland Switch}

A slightly more complicated network is the Goldbeter-Koshland
mechanism. This consists of two coupled
Michaelis-Menten equations. The mechanism is important
because it plays the role of a switch, allowing the propagation
of information in signal transduction networks.
The detailed mechanism is represented by four mass action reactions

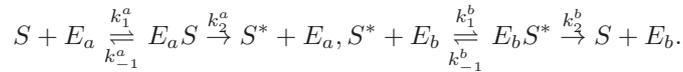
\begin{equation}
S + E_a \underset{k_{-1}^a}{ \overset{k_{1}^a}{\rightleftharpoons}} E_aS  \overset{k_2^a}{\rightarrow}
S^* + E_a,
S^*  + E_b \underset{k_{-1}^b}{ \overset{k_{1}^b}{\rightleftharpoons}} E_bS^*  \overset{k_2^b}{\rightarrow}
S + E_b.
\notag
\end{equation}
where $S$ and $S^*$ are, for instance, the un-phosphorylated
and phosphorylated forms of a substrate, $E_a$, $E_b$, are
kinase and phosphatase enzymes, respectively.

This mechanism leads to the following system of differential equations:
\begin{eqnarray}
 x_1' & = & k_{2}^a x_5 - k_{1}^a x_1 x_3, \notag \\
 x_2' & = & k_{2}^b x_6 - k_{1}^b x_2 x_4, \notag \\
 x_3' & = & k_{-1}^a x_5 + k_{2}^b x_6 - k_{1}^a x_1 x_3, \notag \\
 x_4' & = & k_{2}^a x_5 + k_{-1}^b x_6 - k_{1}^b x_2 x_4, \notag \\
 x_5' & =  & k_{1}^a x_1 x_3 - (k_{-1}^a + k_{2}^a) x_5 , \notag \\
   x_6' & = & k_{1}^b x_2 x_4 - (k_{-1}^b + k_{2}^b) x_6.\label{eq:gkode}
\end{eqnarray}
where $x_1=[E_a]$, $x_2=[E_b]$, $x_3=[S]$, $x_4=[S^*]$,
$x_5=[E_aS]$, $x_6=[E_bS^*]$.

This system has three conservation laws:
\begin{eqnarray}
 &x_1 + x_5 = E_0^a, \notag \\
 &x_2 + x_6 = E_0^b, \notag \\
 &x_3 + x_4 + x_5 + x_6 = S_0.\label{eq:gkconserv}
\end{eqnarray}
Equilibrating each equation of \eqref{eq:gkode} and taking into account
\eqref{eq:gkconserv} leads to the following tropical equations:
\begin{eqnarray}
\gamma_{2}^a \otimes a_5 = \gamma_{1}^a \otimes a_1 \otimes a_3, \notag \\
\gamma_{2}^b \otimes a_6 = \gamma_{1}^b \otimes a_2 \otimes a_4, \notag \\
(\gamma_{-1}^a \otimes a_5) \oplus (\gamma_{2}^b \otimes a_6) = \gamma_{1}^a \otimes a_1 \otimes a_3, \notag \\
(\gamma_{2}^a \otimes a_5) \oplus (\gamma_{-1}^b \otimes a_6) =
\gamma_{1}^b \otimes a_2 \otimes a_4, \notag \\
\gamma_{1}^a \otimes a_1 \otimes a_3 = (\gamma_{-1}^a \oplus \gamma_{2}^a) \otimes a_5 , \notag \\
\gamma_{1}^b \otimes a_2 \otimes a_4 = (\gamma_{-1}^b \oplus \gamma_{2}^b) \otimes a_6, \notag \\
a_1 \oplus a_5 = \gamma_e^a, \notag \\
a_2 \oplus a_6 = \gamma_e^b, \notag \\
a_3 \oplus a_4 \oplus a_5 \oplus a_6 = \gamma_s.\label{eq:tropicalGK}
\end{eqnarray}

The corresponding CSP, described in the next section, is solved instantly and
gives the unique solution: $a_1 = 5, a_2 = 4, a_3 = 3, a_4 = 4, a_5 = 7$ for
parameter values consistent with the literature: $k_1^* = 1000, k_2^* = 150,
k_{-1}^* = 150$.


\section{Tropical Equilibration as a Constraint Satisfaction Problem}

Given a biochemical reaction system with
its Mass-Action kinetics, and a small $\epsilon$, the problem of tropical equilibration is to look for a
rescaling of the variables such that the dominating positive and negative term
in each ODE \emph{equilibrate} as per Definition~\ref{eqreactions}, i.e., are
of the same degree in $\epsilon$.

Note that there are supplementary constraints related to this rescaling when
some conservation laws exist for the original system. Finding these
conservation laws is another CSP which can be solved efficiently with constraint methods \cite{Soliman12amb}.
Here we will assume that the conservation laws are given in input. In our prototype implementation, both the
computation of conservation laws and the following equilibration are performed
for a given system.


For each original equation $dx_i/dt$, $1\leq i\leq n$ is introduced a variable
$a_i\in\mathbb{Z}$ that is used to rescale the system by posing $x_i =
\epsilon^{a_i}\bar{x_i}$. These are the variables of our CSP\@. Note that they
require a solver handling $\mathbb{Z}$ like for instance
SWI-Prolog~\cite{WSTL12tplp,swi} with the \texttt{clpfd} library by Markus
Triska.

The constraints are of two kinds. For each differential equation that should
be equilibrated is a list of positive monomials $M^+_i$, and a list of
negative monomials $M^-_i$. The degrees in $\epsilon$ of all these monomials
are integer linear expressions in the $a_i$. Now, to obtain an equilibration
one should enforce for each $i$ that the minimum degree in $M^+_i$ is equal to
the minimum degree in $M^-_i$. This will ensure that we find two monomials
(\emph{i} of Definition~\ref{eqreactions}) of opposite sign (\emph{ii}) and of
minimal degree (\emph{iii}). This corresponds to the first six tropical
equations of~\eqref{eq:tropicalGK}.
We will see how they can be implemented with reified constraints, but for now, let
us assume a constraint \lstinline|min(L, M)| that enforces that the FD
variable \lstinline|M| is the minimum value of a list
\lstinline|L| of linear expressions over FD variables.
We have in our CSP, for each $1\leq i\leq n$,
\lstinline|min(PositiveMonomialDegrees, M)| and
\lstinline|min(NegativeMonomialDegrees, M)|.

The second kind of constraint comes from conservation laws. Each conservation
law is an equality between a linear combination of the $x_i$ and a constant
$c_i$. By rescaling, we obtain a sum of rescaled monomials equal to
$\epsilon^{\log(c_i)/\log(\epsilon)}\bar{c_i}$. We want this equality to hold
when $\epsilon$ goes to zero, which implies that the minimal degree in
$\epsilon$ in the left hand side is equal to (the round of) the degree of the
right hand side. Since once again the degrees on the left are linear
combinations of our variables $a_i$, this is again a constraint of the form:
\lstinline|min(ConservationLawDegrees, K)| where \lstinline|K| is equal to
$\mathrm{round}(\log(c_i)/\log(\epsilon))$. This corresponds to the last three
tropical equations of~\eqref{eq:tropicalGK}.

Furthermore, if the system under study is not at steady state, the minimum
degree should not be reached only once, which would lead to a constant value
for the corresponding variable when $\epsilon$ goes to zero, but at least
twice. This is the case for the example treated in~\cite{NGVR12sasb}.
The constraint we need is therefore slightly more general than
\lstinline|min/2|: we need the constraint \lstinline|min(L, M, N)| which is true if
\lstinline|M| is smaller than each element of \lstinline|L| and equal to
\lstinline|N| elements of that list.
Note that using CLP notation, we have:
\begin{lstlisting}
min(M, L) :-  C#>=1, min(M, L, C).
\end{lstlisting}

In order to enforce that the minimum is reached at least a required
number of times, one obvious solution is to try all pairs of positive and
negative monomials and count the successful pairs \cite{RGZN12fbcb}.
However, this is not necessary, the \lstinline|min(L, M, N)| constraint
directly expresses the cardinality constraint on the minimums.
and can be implemented using {\em reified constraints} to propagate information
between \lstinline|L|, \lstinline|M| and \lstinline|N| in all directions,
without enumeration.
Using SWI-Prolog notations, the implementation of \lstinline|min/3|
by reified constraints is as follows:
\begin{lstlisting}[language=Prolog]
min([], _, 0).
min([H | T], M, C) :- M#=<H, B #<==> M#=H, C#=B+CC,
                      min(T, M, CC).
\end{lstlisting}

This concise and portable implementation will probably improve when the
\lstinline|minimum| and \lstinline|min_n| global constraints are available
(see~\cite{BCDP05sics} for a reference). However 
it
already proves very efficient as demonstrated in the next section.

\section{Computation Results on Biomodels.net}

To benchmark our approach, we applied it systematically to all the dynamical models of
the BioModels\footnote{\texttt{http://biomodels.net}}
repository~\cite{NBBCDDLSSSSH06nar} of biological systems,
with $\epsilon$ set arbitrarily to $0.1$.
We used the latest release (\emph{r24} from 2012-12-12) which includes 436
curated models.

Among them, only 55 models have non-trivial purely polynomial kinetics
(ignoring \emph{events} if any). Our computational results on those are
summarized in the following table, where
the first column indicates whether a complete equilibration was found, and the
times are in seconds.

\begin{center}
   \begin{tabular}{|r|r|r|r|}
      \hline
      Found & \# models & Variables (avg/min/max) & Time (avg/min/max) \\
      \hline
      yes & 23 &  17.348/3/$\;\,$86 & 0.486/0.004/2.803 \\
      \hline
      no & 32 &  17.812/1/194 & 0.099/0.000/1.934 \\
      \hline
   \end{tabular}
\end{center}

We managed to avoid timeouts by using an iterative domain expansion: the
problem is first tried with a domain of $[-2,2]$, i.e., equilibrations are
searched by rescaling in the $10^{-2}, 10^2$ interval. If that fails, the
domain is doubled and the problem tried again (until a limit of $10^{-128},
10^{128}$). This strategy coupled with a domain bisection enumeration
(\lstinline|bisect| option in SWI-Prolog) allowed us to gain two orders of
magnitude on the biggest models.

Only one of the models (number 002) used values far from 0 in the
equilibration (up to $\epsilon^{40}$) and has no complete equilibration if the
domain is restricted to $[-32,32]$. This is because all kinetics are scaled by
the volume of the compartment, which in that case was $10^{-16}$, translating
the search accordingly. We thus do not believe that enlarging the domains even
more would lead to more equilibrations. Nevertheless, choosing a smaller
$\epsilon$ might increase the number of equilibrations.

18 of the 23 models for which there is a complete equilibration are actually
underconstrained and appear to have an infinity of such solutions (typically
linear relations between variables). For the 5 remaining ones, we computed all
complete equilibrations:

\begin{center}
   \begin{tabular}{|r|r|r|}
      \hline
      Model & \# equilibrations & Total time (s) \\
      \hline
      BIOMD0000000002 & 36 & 109\\
      \hline
      BIOMD0000000122 & 45 & 291\\
      \hline
      BIOMD0000000156 & 7  & 0.008\\
      \hline
      BIOMD0000000229 & 7  & 0.7\\
      \hline
      BIOMD0000000413 & 29 & 3.3\\
      \hline
   \end{tabular}
\end{center}

\section{Discussion}
\label{conclusion}

One of the limits of this approach, is that it is not well suited to
equilibration problems with an infinite number of solutions.
\corr{For those, symbolic solutions depending on free parameters are necessary, as done in~\cite{NGVR13arxiv}.}

It is also possible to reduce a system using its conservation laws, and to apply
tropical equilibration directly on the reduced system. However, the resulting equilibrations
might be slightly different, apparently due to the possible loss of
positivity of certain variables. We want to investigate this question further.

In many cases, it makes sense biologically to only look for partial
equilibrations. Strategies to decide when such decision has to be made remain
unclear. Nevertheless the framework of partial constraint satisfaction and
more specifically Max-CSP~\cite{FW92ai} would allow us to easily handle the
maximization of the number of equilibrated variables.

\corr{
In this paper we discussed only the calculation of the tropical equilibrations and
of the unknown orders of the variables. Once the orders of the variables are known, the
rapid variables can be identified and the system reduced to a simpler one. The details
of the reduction procedure, involving pruning of dominated terms  and pooling
 of fast variables into fast cycles  will be presented elsewhere. A simple
reduction procedure, involving only pruning is described by
Theorem 3.6 of \cite{NGVR13arxiv}.}

\subsubsection*{Acknowledgements.}
This work has been supported by the French ANR Bio\-Tempo, \corr{CNRS Peps ModRedBio,} and OSEO Biointelligence projects.


\begin{thebibliography}{}

\end{thebibliography}


\begin{thebibliography}{10}

\bibitem{BCDP05sics}
N.~Beldiceanu, M.~Carlsson, S.~Demassey, and T.~Petit.
\newblock Global constraints catalog.
\newblock Technical Report T2005-6, Swedish Institute of Computer Science,
  2005.

\bibitem{CGQ99arc}
G.~Cohen, S.~Gaubert, and J.P. Quadrat.
\newblock Max-plus algebra and system theory: where we are and where to go now.
\newblock {\em Annual Reviews in Control}, 23:207--219, 1999.

\bibitem{FW92ai}
Eugene~C Freuder and Richard~J Wallace.
\newblock Partial constraint satisfaction.
\newblock {\em Artificial Intelligence}, 58:21--70, 1992.

\bibitem{NBBCDDLSSSSH06nar}
Nicolas le~Nov{\`e}re, Benjamin Bornstein, Alexander Broicher, M{\' e}lanie
  Courtot, Marco Donizelli, Harish Dharuri, Lu~Li, Herbert Sauro, Maria
  Schilstra, Bruce Shapiro, Jacky~L. Snoep, and Michael Hucka.
\newblock {BioModels Database}: a free, centralized database of curated,
  published, quantitative kinetic models of biochemical and cellular systems.
\newblock {\em Nucleic Acid Research}, 1(34):D689--D691, January 2006.

\bibitem{NGVR12sasb}
Vincent Noel, Dima Grigoriev, Sergei Vakulenko, and Ovidiu Radulescu.
\newblock Tropical geometries and dynamics of biochemical networks application
  to hybrid cell cycle models.
\newblock In J{\'e}r{\^o}me Feret and Andre Levchenko, editors, {\em
  Proceedings of the 2nd International Workshop on Static Analysis and Systems
  Biology ({SASB} 2011)}, volume 284 of {\em Electronic Notes in Theoretical
  Computer Science}, pages 75--91. Elsevier, 2012.

\bibitem{NGVR13arxiv}
Vincent Noel, Dima Grigoriev, Sergei Vakulenko, and Ovidiu Radulescu.
\newblock Tropicalization and tropical equilibration of chemical reactions.
\newblock arXiv:1303.3963, in press Contemporary Mathematics, 2013.

\bibitem{RGZN12fbcb}
O.~Radulescu, A.N. Gorban, A.~Zinovyev, and V.~Noel.
\newblock {Reduction of dynamical biochemical reaction networks in
  computational biology}.
\newblock {\em Frontiers in Bioinformatics and Computational Biology}, 3:131,
  2012.

\bibitem{Soliman12amb}
Sylvain Soliman.
\newblock Invariants and other structural properties of biochemical models as a
  constraint satisfaction problem.
\newblock {\em Algorithms for Molecular Biology}, 7(15), May 2012.

\bibitem{Viro08jjm}
O.~Viro.
\newblock From the sixteenth {H}ilbert problem to tropical geometry.
\newblock {\em Japanese Journal of Mathematics}, 3(2):185--214, 2008.

\bibitem{swi}
Jan Wielemaker.
\newblock {\em {SWI-Prolog} 6.3.15 Reference Manual}, 1990--2013.

\bibitem{WSTL12tplp}
Jan Wielemaker, Tom Schrijvers, Markus Triska, and Torbj\"o{}rn Lager.
\newblock {SWI-Prolog}.
\newblock {\em Theory and Practice of Logic Programming}, 12(1-2):67--96, 2012.

\end{thebibliography}
\end{document}